\documentclass[times,final,authoryear]{elsarticle}
\usepackage{graphicx}
\usepackage{rotating}
\usepackage{epstopdf}
\usepackage{epsf}
\usepackage{longtable}
\usepackage{savesym}
\savesymbol{tablenum}
\usepackage{siunitx}
\restoresymbol{SIX}{tablenum}
\usepackage{amssymb}
\usepackage{amsmath}
\usepackage{gensymb}
\usepackage{latexsym}
\usepackage{wrapfig}
\usepackage{afterpage}
\usepackage{xcolor}
\usepackage{color, colortbl}
\usepackage{comment}
\usepackage{threeparttable}
\usepackage[switch]{lineno}

\usepackage{url}

\expandafter\ifx\csname package@font\endcsname\relax\else
 \expandafter\expandafter
 \expandafter\usepackage
 \expandafter\expandafter
 \expandafter{\csname package@font\endcsname}

\fi
\def\bq{\begin{equation}}
\def\eq{\end{equation}}
\def\bqy{\begin{eqnarray}}
\def\eqy{\end{eqnarray}}

\newcommand{\hl}[1]{\textbf{\large{#1}}}
\graphicspath{{./}{figures/}}
\usepackage{jasr}
\usepackage{framed,multirow}

\usepackage{url}
\usepackage{xcolor}
\definecolor{newcolor}{rgb}{.8,.349,.1}

\usepackage[citebordercolor=white]{hyperref}

\journal{Advances in Space Research}
\begin{document}
\verso{Given-name Surname \textit{etal}}
\begin{frontmatter}
\title{Project Lyra: The Way to Go and the Launcher to Get There }%

\author[1]{Adam \snm{Hibberd}\corref{cor1}}
\cortext[cor1]{Corresponding author: 
Tel.: +0-000-000-0000;
fax: +0-000-000-0000;}
\ead{adam.hibberd@i4is.org}


\affiliation[1]{organization={Initiative for Interstellar Studies (i4is)},
addressline={27/29 South Lambeth Road},
city={London},
postcode={SW8 1SZ},
country={United Kingdom}}

\begin{abstract}
In preceding papers, Project Lyra has covered many possible trajectory options available to a spacecraft bound for 1I/'Oumuamua, including Solar Oberth manoeuvres, Passive Jupiter encounters, Jupiter Oberths, Double Jupiter Gravitational Assists, etc. Because \emph{feasibility} was the key driver for this analysis, the important question of which launcher to exploit was largely skirted in favour of adopting the most powerful options as being \emph{sufficient}, though these launchers are clearly not \emph{necessary}, there being alternative less capable candidates which could be utilised instead. In this paper the various launch options available to Project Lyra are addressed to allow a general overview of their capabilities. It is found that the SpaceX Super-Heavy Starship would be a game-changer for Project Lyra, especially in the context of refuelling in LEO, and furthermore a SpaceX Falcon Heavy Expendable could also be utilised. Other launchers are considered, including Ariane 6 and the future Chinese Long March 9. The importance of the $V_{\infty}$ Leveraging Manoeuvre (VILM) in permitting less capable launchers to nevertheless deliver a payload to 'Oumuamua is elaborated.
\end{abstract}

\end{frontmatter}

\section{Introduction}
\label{sec1}
Project Lyra is the feasibility study of missions to the first interstellar object to be discovered, 1I/’Oumuamua \citep{Flekky2019}.
The Lyra papers \citep{HEIN2019552,Hibberd_2020,HIBBERD2021,Hibberd_2022,HIBHEIN,hibberd2022project,hibberd2023project} have addressed various possible trajectory options but have largely skirted the issue of which launcher should carry the Lyra spacecraft payload by assuming one or two select launchers. The NASA Space Launch System Block 2 is the outstanding candidate for such a mission due to its \emph{super-heavy} lift credentials.\\

Broadly speaking there are three possible trajectory options available to a putative craft destined for ‘Oumuamua:

\begin{enumerate}
\item{The Solar Oberth Manoeuvre (SOM)}
\item{The Jupiter Oberth Manoeuvre (JOM)}
\item{The Passive Jupiter Gravitational Assist (PJGA)}
\end{enumerate}

Since ‘Oumuamua is on a hyperbolic orbit (by definition), has passed perihelion, and is now heading away in excess of 26 \si{km.s^{-1}} (equivalent to 5.5 \si{au.year^{-1}}), an intercept mission must seek ways to rapidly escape the Solar System with a heliocentric excess speed of at least this value, in order to catch it up. 
As elaborated in \cite{Blanco21}, the SOM above is optimal for achieving this requirement, that is if one ignores practicalities such as the necessity for a heat shield to protect the craft from the powerful solar flux associated with the close slingshot of the Sun. In addition the SOM option requires a trip to Jupiter in order to do a gravity brake and achieve a close perihelion to the Sun. The early Lyra papers naturally adopted the SOM as the optimal strategy for travelling to ‘Oumuamua, and so therefore factored in a carbon-carbon composite heat shield (scaled up from the Parker Solar Probe’s). The results using this option were relatively rapid intercept of the target ($\sim{22}$ \si{years}) but also have a rather lower technological maturity level (TRL) due to the innovative SOM requirement, a manoeuvre never so far implemented in any mission, and would require rapid technological advancement in the field.\\

\begin{table}
\centering
\caption{Possible Launch Vehicles for Project Lyra.}
\label{Launchers}
\begin{threeparttable}
\begin{tabular}{|c|cccccc|c|} \hline
 & \textbf{Orbit $\rightarrow$} & \textbf{LEO} &\textbf{GTO} & \textbf{Lunar} & \textbf{Mars} & \textbf{Jupiter} & \textbf{Units}\\
\textbf{Launch Vehicle $\downarrow$} & \textbf{$C_3$ $\rightarrow$} & \textbf{-60} & \textbf{-16.3} & \textbf{0} & \textbf{12} & \textbf{84} & \textbf{\si{km^{2}.s^{-2}}}\\ \hline
Ariane 5\tnote{a,c} &Heavy Lift&20&9.2&6.6 \tnote{d} &4.1&N/A&mt\\
Ariane 6 4\tnote{b,e} & Heavy Lift & 21.65 & 11.5 & 8.6 & 6.9 & N/A & mt\\
Delta IV Heavy\tnote{a,f}& Heavy Lift & 28.79& 14.22 & 10 & 8\tnote{g} & N/A & mt\\
Falcon Heavy Exp.\tnote{h} &	Heavy Lift & ? & ? & 15.01 & 11.88 & 1.875 & mt \\
SH + Starship\tnote{b,i}  &Super-Heavy Lift & 150 & 21	& ?	& ?	& ? & mt \\
Long March 9\tnote{b,j}  & Super-Heavy Lift	& 150 & ? &	54 & 44	& ? & mt \\
SLS Block 2\tnote{b,k} & Super-Heavy Lift & 130 & 58 & 46 & 37 & 8 & mt\\ \hline
\end{tabular}
\begin{tablenotes}
\item [a] Launchers to be Phased out Soon
\item [b] Imminently Available Launch Capabilities
\item [c] \cite{Ariane5}
\item [d] Sun/Earth Lagrange 2 Point
\item [e] \cite{Ariane6}
\item [f] \cite{DeltaIV}
\item [g] \cite{DeltaIV_2}
\item [h] from NASA Launcher Query Service \citep{NASA_Q}
\item [i] \cite{Starship}
\item [j] \cite{LongMarch}
\item [k] \cite{SLS}
\end{tablenotes}
\end{threeparttable}

\end{table}
In this document the SOM is excluded from the analysis, largely because reasonably effective missions using either the JOM or the PJGA can be realised by the powerful launchers to be considered herein. Furthermore the PJGA and the JOM are, in that order, the priorities defined in the Interstellar Probe Project as is stated in the definition file produced by JHU APL \citep{InterstellarProbe}. \\

As far as the selection of a launcher is concerned, the reason for the narrow line of enquiry until now, is the abundance of information on the SLS made available by NASA. Moreover, in specific regard to escape missions – i.e. missions which escape the gravitational influence of our home world in order to travel to distant destinations in our Solar System - many launch vehicle user guides, especially future ones, give only a broad estimate of the capability of the vehicle in question. \\

Herein we shall attempt to redress this bias by exploring the possibilities of using the many alternative launch systems which may be available in the Project Lyra launch timeline (i.e. 2026-2033), some of which are currently operational, and some which soon will be.\\

\begin{figure}
\hspace{-0.4cm}
\includegraphics[scale=0.42]{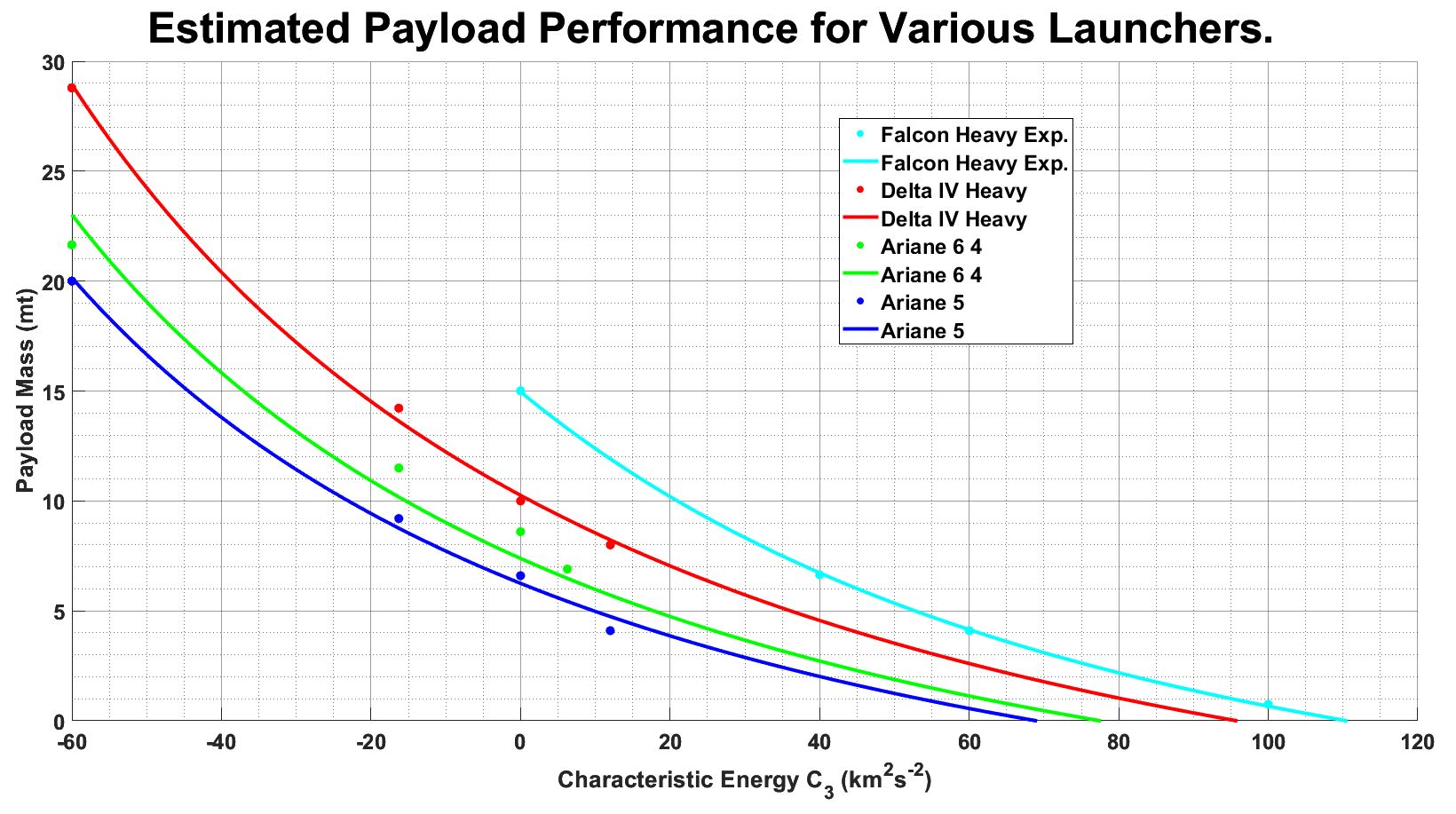}
\caption{Escape Mission Performance of Launchers Considered Herein. These are Estimates Only}
\label{fig:Perf}
 \end{figure}

The main candidates are outlined in Table \ref{Launchers}, Launchers corresponding to capabilities which will expire in the near future are labelled with the letter 'a' and so will not be treated in the following analysis. Those labelled with the letter 'b' will, in principle, be available by the time of Project Lyra. Falcon Heavy Expendable is currently operating and is planned to continue to do so in the foreseeable future.\\

The performance statistics provided in Table 1 are insufficient to conduct any useful research. Thus from these values a graph of performance data for the first four launchers listed in Table 1 was constructed, in order to incorporate a broader range of $C_3$values. Refer to Figure \ref{fig:Perf}. The solid lines indicate best fits of the data (provided by the circular dots) and allow some sort of extrapolation to higher $C_3$ values, especially for Ariane 5 and Ariane 6 4. It must be emphasised that this graph required some degree of guess-work.\\

\section{Ariane 5 \& 6}

Launched near the equator (which is optimal for GTO missions) from a location at Kourou, French Guyana, the Ariane family of launchers is the venerable workhorse of the European space program. Ariane 5 is operated by Arianespace, who are looking to phase it out in the next few years, in favour of the more powerful Ariane 6 (though not so powerful as the super-heavy launchers referred to later in this note).\\

Ariane 6 4 will have 4-strap on boosters, making it a considerable step-up from the Ariane 6 2 (2 strap-on boosters) which should be launched later this year.\\

Referring to Table \ref{Launchers}, for the ESA launch vehicles Ariane 5 \& Ariane 6 4, they appear to have insufficient performance to reach Jupiter directly, and as you may have observed from the various Project Lyra literature, this is an important pre-requisite to undertake any sort of mission to ‘Oumuamua. \\

Due to it soon being retired from service, we must reject Ariane 5 as a candidate. This leaves Ariane 6 4. From Table \ref{Launchers} Ariane 6 4 has a capability to $C_3 = 0\ \si{km^2.s^{-2}}$ of 8.6 \si{mt}. Clearly this $C_3$ is far too low to make any direct journey to Jupiter, in fact it is barely sufficient to escape the Earth’s gravitational pull. \\

However there is a possible route using this $C_3$, which might enable some sort of feasible mission for this heavy lift launcher, and that is the \emph{$V_{\infty}$ Leveraging Manoeuvre} (VILM) \cite{SimsLong} \\

For a VILM, the spacecraft (s/c) embarks on an Earth-return heliocentric elliptical arc, with a time-period of n multiples of Earth’s year (365 days), where n is a whole number, usually 1, 2 or 3. A VILM is a useful mechanism by which the speed of the s/c relative to the Sun can be augmented via exploiting the Earth’s mass with a gravitational assist (GA) of the planet. The hyperbolic excess relative to the Earth, $V_{\infty}$, required for the n=1 option (i.e. in 1:1 resonance with Earth) is precisely zero and since $C_3 = V_{\infty}^2$, that means $C_3$ is also zero, as is required. As a note, an additional $\Delta V$ of 0.5 \si{km.s^{-1}} is factored in for the Deep Space Manoeuvre (DSM) at 1.0 \si{au} between launch and the return to Earth, so that a high specific impulse electric propulsion can deliver this $\Delta V$ over the course of the resonant orbit as an alternative to a chemical rocket.\\

So we have 8,600kg of capacity for Ariane 6 4, if we introduce two further boosters ‘into the equation’, the STAR 63F and the STAR 48B - refer Table \ref{Boosters} - then that gives us a total of 6,727kg, well within the scope of an Ariane 6 4. Let us further assume a s/c payload mass of 100kg, yielding a total of 6,827kg (there is still capacity for more, but this shall do for the moment). \\

Applying the Tsiolkovsky rocket equation using these two stages, we get a total $\Delta V$ generated by this pair of boosters as 9.30 \si{km.s^{-1}}. It is best to apply ALL this $\Delta V$ at the Earth return, and NONE at Jupiter itself, thence utilising a passive GA at Jupiter to eventually arrive at the target, ‘Oumuamua. (To reinforce this note that the STAR 63F would only be able to deliver 2.85 \si{km.s^{-1}} of $\Delta V$ to a STAR 48B + 100 kg payload, which would be quite insufficient to reach Jupiter – thus instead both stages need to be fired at Earth return.) \\

\begin{table}
\centering

\begin{tabular}{|cccccc|} \hline
Booster Stage &	Exhaust Velocity &	Total Mass & Dry Mass & Propellant Mass & Height \\ \hline
STAR 75	& 2.8225 &	8068 &	565 &	7503 &	2.59\\
STAR 63F &	2.9106 & 4590 &	326	 & 4264 & 1.78\\STAR 48B &	2.8028 & 2137 & 124	& 2013 & 2.03\\
ORION 50XL & 2.8647 & 4306 & 367 & 3939 & 3.07\\
CASTOR 30B & 2.9649	& 13971 & 1000 & 12971 & 3.5\\
CASTOR 30XL & 2.8866 & 26406 & 1392	& 25014	& 6.0\\ \hline
\end{tabular}
\caption{Possible Solid Boosters Available for Project Lyra}
\label{Boosters}
\end{table}
Executing \emph{Optimum Interplanetary Trajectory Software} (OITS), \cite{OITS_info}, with these parameters, combined with the E-DSM-E-J-1I series of encounters, results in a viable mission to the interstellar object in question. \\

\begin{figure}
\includegraphics[scale=0.76]{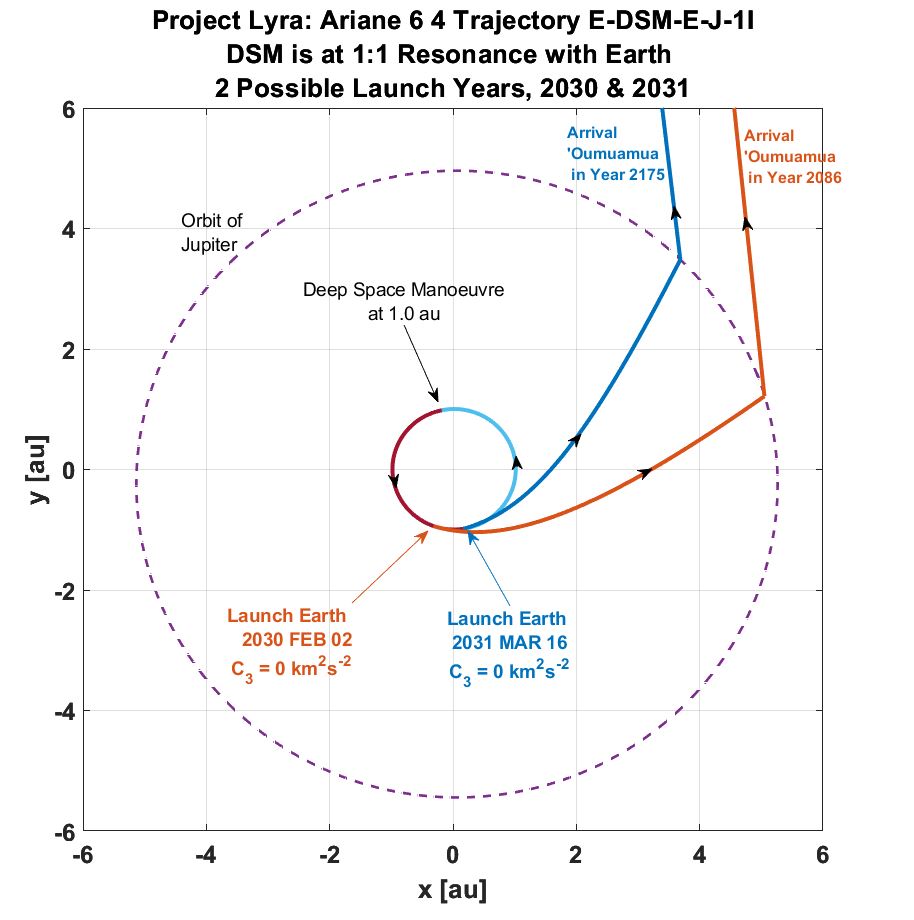}
\caption{Ariane 6 4 Trajectory Options Utilising a VILM in 1:1 resonance with Earth, Launch Years 2030 \& 2031}
\label{fig:Ariane6}
 \end{figure}

Thus we find that there are two candidate years for launch with an Ariane 6 4, both of which involve a hugely protracted mission duration, well in excess of 50 years – refer Figure \ref{fig:Ariane6}. \\

All this effectively rules out the Ariane 6 4 launch vehicle for missions to ‘Oumuamua, unless some alternative, hugely more effective propulsion system can be harnessed in place of the solid propellant stages adopted here.

\section{Delta IV Heavy}
United Launch Alliance (ULA) will retire this launch vehicle in 2024, so this launcher can be neglected for Project Lyra.

\section{Falcon Heavy Expendable}
The SpaceX Falcon Heavy is a super-heavy launch vehicle which at the time of writing, has executed five successful launches. The fifth most powerful launcher of all time, this launch vehicle has two configurations: either partially reusable or expendable, the latter more capable version is investigated here.\\

\begin{figure}
\hspace{-1.2cm}
\includegraphics[scale=0.45]{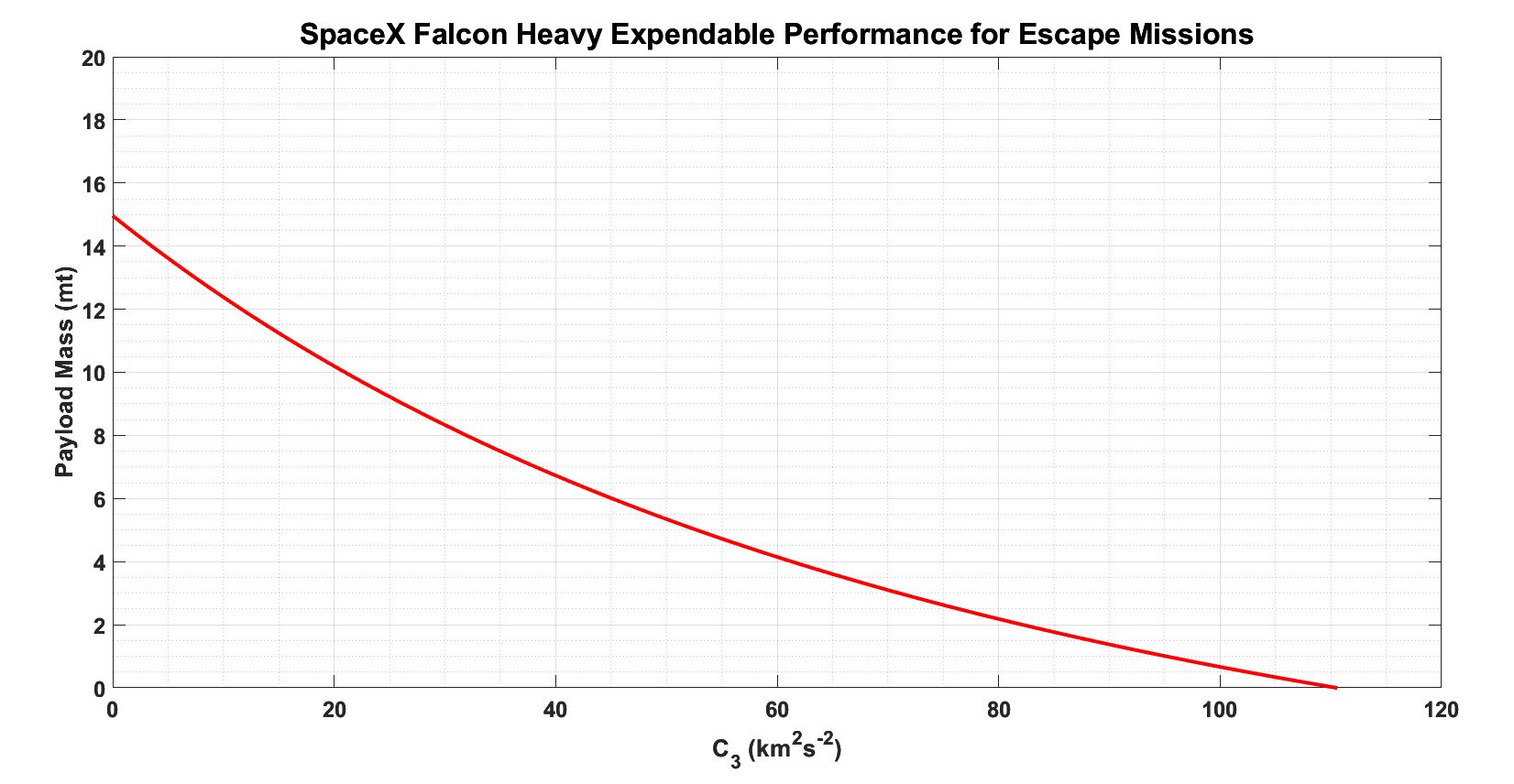}
\caption{Falcon Heavy Expendable Performance for Escape Missions}
\label{fig:FH_Perf}
 \end{figure}
From Figure \ref{fig:FH_Perf}, a direct mission to Jupiter is feasible using this SpaceX rocket, the drawback being however that a payload mass less than 3000kg is indicated. From Figure \ref{fig:FH_Perf}, we find that a $C_{3}$ value of 100 $\si{km^2.s^{-2}}$, which gets to Jupiter and more, can deliver a mass of 750kg to the escape orbit from Earth. Is this sufficient to allow a mission to ‘Oumuamua within realistic timeframe? The answer is a definite no. \\ 

Let us instead settle on a VILM, as we did with Ariane 6 4 but, with a far more powerful capability than the Ariane 6 4 could afford, we have a choice of n=1, 2 or 3, and the $C_3$ for each option is presented in the Table \ref{VILM_FH}.\\

\begin{table}

\centering
\begin{tabular}{|c|c|c|c|c|c|} \hline
n & Time Period (days) & Aphelion (au) & $V_{\infty}$ (\si{km.s^{-1}}) & $C_3$ (\si{km^2.s^{-2}}) & FH Mass (\si{mt})\\ \hline
1 &	365	& 1.0 &	0 & 0 &	15 \\
2 & 730 & 2.2 & 5.140 & 26.42 & 10 \\
3 & 1095 & 3.2 & 6.982 & 48.7 & 6 \\ \hline
\end{tabular}
\caption{VILM Requirements for Falcon Heavy Missions to 'Oumuamua}
\label{VILM_FH}
\end{table}
Table \ref{Results_FH} provides the results of this analysis, addressing each of the scenarios n=1, 2 \& 3 in turn. As can be observed, the option n=2, hits the \emph{sweet-spot} for this Falcon Heavy launcher, with an overall flight duration of 28 years, and with the n=1 \& 3 lagging quite significantly behind, at 54 years and 43 years respectively.\\

\begin{table}[]
    \centering
    \begin{tabular}{|c|c|c|c|c|c|c|c|} \hline
      n &	Mass avail.  & Stages &	Mass used + & $\Delta V$ at Earth & 
$\Delta V$ at &	Flight & Launch Date \\ 
      & (mt) & & Payload 100kg & Return & Jupiter & Duration & \\
    &  & (mt) & (\si{km.s^{-1}}) & (\si{km.s^{-1}}) & (years) & & \\ \hline
1 &	15 & STAR 75 + STAR 48B	& 10,305 & 10.126 &	0 &	54 &	2029 DEC 19 \\
2 &	10 & STAR 63F + STAR 48B x2	& 9,857 & 3.6074 & 6.4499 &	28 & 2029 APR 08 \\
3 &	6 &	STAR 48B x2 & 4,374 & 1.7282 & 6.4499 &	43 & 2027 FEB 25 \\ \hline
\end{tabular}
\caption{Results of Falcon Heavy Analysis with VILM Missions}
\label{Results_FH}
\end{table}

The n=2 scenario supposes that a STAR 63F and a STAR 48B are fired at the Earth return, and then a second STAR 48B is fired at perijove. Further, as in the case of Ariane 6 4, an additional 0.5 \si{km.s^{-1}} is applied at the DSM, the reason for adopting this value here is so that, as an alternative to chemical, an electric low-thrust propulsion system with high specific impulse would, over the course of the resonant orbit, be able to apply this $\Delta V$, as required.\\

For an animation of the n=2 solution go to \cite{FH_anim}.\\\

The declination of the escape asymptote for the n=2 option is $-23.0\degree$ and its magnitude is marginally less than the launch latitude for the Kennedy Space Center, LC-39A (latitude $\sim{28}\degree$). This means that the Falcon Heavy would potentially be able to launch on an (optimal) eastwards azimuth.

\section{Super-Heavy Starship} 

The SpaceX Starship will be a hugely capable launch vehicle, and there has been at the time of writing, one attempt to orbit which was a failure after first stage burnout. The specification data on the internet for this launch vehicle is scant, but there is the User Guide \citep{SpaceX1}.\\

With a good deal of reference to source documents on the internet, Table \ref{Comp_L} was constructed. The plain text numbers are data garnered from the internet, whilst the larger font, highlighted numbers are data derived from the plain font parameters through appropriate calculations. \\

In Table \ref{Comp_L}, the Starship is compared with an SLS Block 1 to see if any inferences can be made from this comparison and in order to get some idea as to the performance characteristics of the Starship.
We find that for an LEO, the SLS Block 1 and the Starship both have around the same calculated total $\Delta V$, to around 0.05\%.\\

\renewcommand{\arraystretch}{1.25}
\afterpage{
\centering

\begin{table}[ht] 
\caption{Comparison of NASA SLS Block 1 Launch Vehicle with the SpaceX Super-Heavy \& Starship\\Normal font are inputs\\Larger font are calculated from inputs }
\label{Comp_L}
\begin{threeparttable}
\begin{tabular}{|l|l|l|l|l|} \hline
& & & & \\
& & \textbf{SLS Block 1} & \textbf{SH \& Starship} & Notes \\
& & & & \\ 
\hline
First Stage &
Ve ($\si{km.s^{-1}}$) & 4.0479 & 3.2046\tnote{g} &  \\ \hline
& Mass Propellant (kg)& 987471.0000\tnote{b}& 3600000.0000 \tnote{g}& \\ \hline
& Dry Mass (kg) & 85275.0000\tnote{b}& 160000.0000\tnote{g}& \\ \hline
& Burn time (s) & 500.0000\tnote{b}& & \\ \hline
 & Mass Flow Rate (\si{kg.s^{-1}}) &\hl{1974.9420} & &  \\ \hline 
& & & & \\ \hline
Strap-on Boosters & Ve (\si{km.s^{-1}})& 2.6360\tnote{c}& &\\ \hline
& Mass Propellant (kg)& 1451496.\tnote{c}& &\\ \hline
& Dry Mass (kg) & 200778.0000\tnote{c} & &  \\ \hline
& Burn Time (s) & 126.0000\tnote{c} & &  \\ \hline
 & Mass Flow Rate (\si{kg.s^{-1}})& \hl{11519.8095}& &\\ \hline
 & & & & \\ \hline
Second Stage& Ve (\si{km.s^{-1}})& 4.5090\tnote{d} & 3.4300 \tnote{g} &  \\ \hline
 &Mass Propellant (kg)& 29000.0000\tnote{d}& 1200000.0000 \tnote{g} &  \\ \hline
& Dry Mass (kg) & 3700.0000\tnote{d} & 100000.0000 \tnote{g}&  \\ \hline
& & & &\\ \hline
LEO & Payload Mass (kg) & 95000.0000\tnote{a} & 150000.0000 \tnote{f} &\\ \hline
& $\Delta V$ Boosters On (\si{km.s^{-1}})& \hl{2.5745}& & \\ \hline
& $\Delta V$ Boosters Ej (\si{km.s^{-1}})& \hl{6.0376}& & \\ \hline
& $\Delta V$ First Stage (\si{km.s^{-1}})& \hl{8.6120}& \hl{3.7633} & \\ \hline
& $\Delta V$ Second Stage (\si{km.s^{-1}}) & \hl{1.2238}& \hl{6.0295} & \\ \hline
& & & & \\ \hline
 &$\Delta V$ Total (\si{km.s^{-1}})& \hl{9.7978}& \hl{9.7928} &  Difference $\sim{0.05}$ \% \\ \hline
& & & & \\ \hline
Trans-Lunar & Payload Mass (kg) & 27000.0000\tnote{e} & \hl{*0.0000*} & *Assuming NO MASS\\ \hline
& $\Delta V$ Boosters On (\si{km.s^{-1}})& \hl{2.6786}& &  To The Moon for\\ \hline
& $\Delta V$ Boosters Ej (\si{km.s^{-1}})& \hl{7.2816}& &  Starship* \\ \hline
& $\Delta V$ First Stage (\si{km.s^{-1}})& \hl{9.9602}& \hl{*3.9831*} &  \\ \hline
& $\Delta V$ Second Stage (\si{km.s^{-1}}) & \hl{3.1416}& \hl{*8.7978*} &  \\ \hline
& & & & \\ \hline
& $\Delta V$ Total (\si{km.s^{-1}})& \hl{13.1018} & \hl{*12.7809*}&  Difference $\sim{2.5}$ \% \\ \hline
& & & & \\ \hline
\end{tabular}
\begin{tablenotes}
\item [a] \cite{NASA1}
\item [b] \cite{NASA2}
\item [c] \cite{NASA3}
\item [d] \cite{NASA4}
\item [e] \cite{NASA5}
\item [f] \cite{SpaceX1}
\item [g] \cite{SpaceX2}
\end{tablenotes}
\end{threeparttable}
\end{table}
}

 \begin{table*}
\centering
\begin{tabular}{|c|c|c|c|c|c|} \hline
Launch Vehicle & Additional& Payload&	Launch Date & Flight & Cost/kg \\
& Booster Stages & Mass (kg) & & Dur. (Years) &\\ \hline
Ariane 6 4 & STAR 63F \& STAR 48B &	100	& 2030 FEB 02 &56	& €3187= \$3417\\
Falcon Heavy & STAR 63F \& 2xSTAR 48B & 100 & 2029 APR 08 & 28 & \$1500 \\ 
Super Heavy \& Starship & C. 30XL \& C. 30B \& STAR 48B &860 & 2031 FEB 28 &	20 & \$150  \\
Long March 9 & STAR 63F \& STAR 48B & 100 & 2030 JAN 13 & 36 & ? \\ \hline
\end{tabular}
\caption{Results Summary}
\label{tab:Summary}
\end{table*}
For a Trans-Lunar Orbit (TLO) , however, where the $C_3$ is around 0 \si{km^2.s^{-2}}, the Starship is found significantly wanting, even with a supposed zero payload mass. Furthermore this would translate to an even larger degree of inadequacy for interplanetary missions. \\

Conclusion: Given the data at hand, the Super Heavy Starship is incapable of delivering a spacecraft directly to an Earth escape orbit. In fact this is precisely what the SpaceX Starship User Guide suggests.\\

Let us instead look at the payload it can deliver to LEO, that amounts to an enormous 150 mt. However, the whole context of the Starship design is to allow the potential for in-space refuelling. Would this important asset permit missions to ‘Oumuamua with much lower flight durations? We find that to entirely refuel a Starship in LEO, would need 8 launches of SH + Starship, each one carrying 150 mt of fuel. \\
\begin{wrapfigure}{l}{0.53\textwidth}
\includegraphics[width=0.56\textwidth]{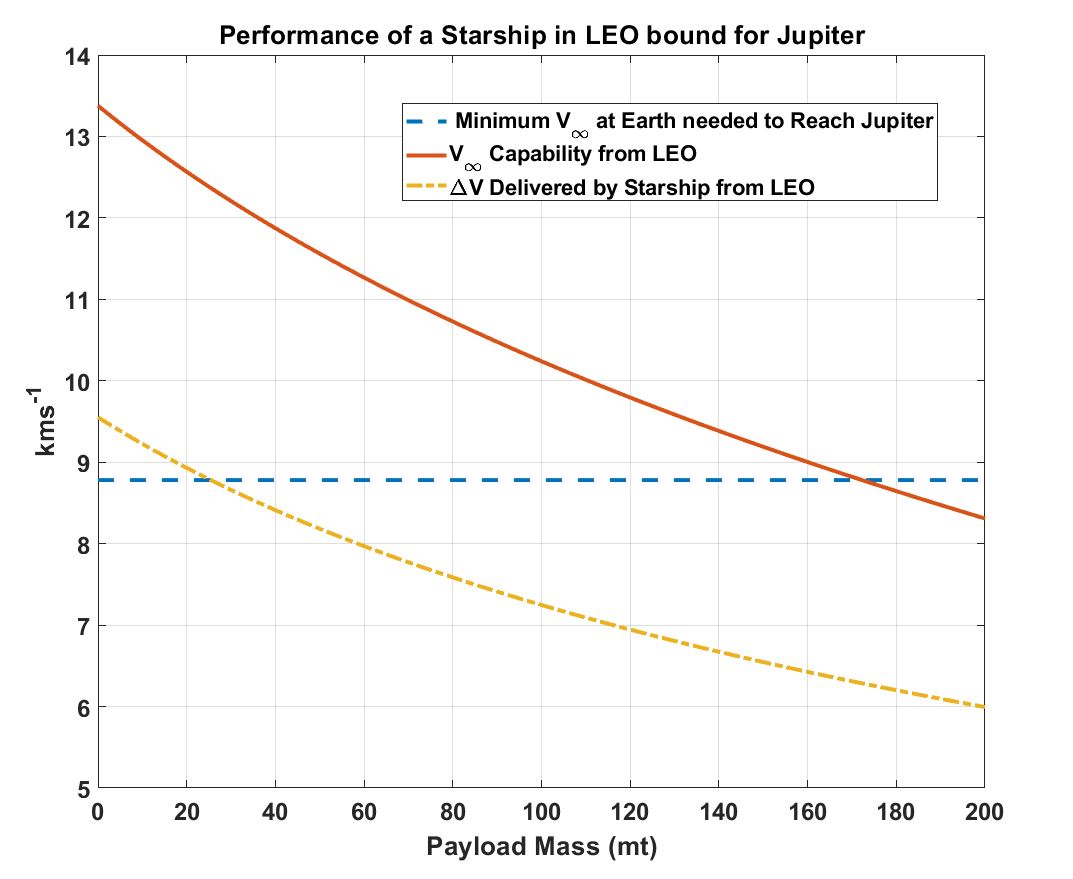}
\caption{Refuelled Starship in LEO Capability to Jupiter}
\label{fig:Star_Perf}
 \end{wrapfigure}

Given that we have a fully refuelled Starship in LEO, what can we do to leverage this asset for a mission to ‘Oumuamua via Jupiter? Figure \ref{fig:Star_Perf} provides the necessary parameters for a Starship to exit LEO and carry a payload to Jupiter. The horizontal blue-dashed line is the minimum hyperbolic excess speed, $V_{\infty}$, from Earth to travel along a Hohmann Transfer to Jupiter (i.e. the theoretical minimum ‘energy’ route to Jupiter). From this line we see that a fuelled Starship in LEO can send any mass up to around 170 mt to Jupiter. 
This single important fact opens a wealth of options from which to choose in terms, for instance, of leveraging some combination of Booster Stages given in Table \ref{Boosters}, to burn at perijove and arriving at ‘Oumuamua extremely rapidly.\\

Let us address two such combination as example cases, \textbf{Scenario 1} is a 2 stage option, and \textbf{Scenario 2} is a 3 stage option (where in both cases the stages are fired at perijove).\\

\textbf{Scenario 1 (2 stages):}
We shall assume a payload mass of 860 kg (the same as the proposed Interstellar Probe -\cite{InterstellarProbe}. Together with a CASTOR 30XL and a STAR 75, this gives a total mass of 35.3 mt, and referring to Figure 3, this leads to a $V_{\infty}$ of around 12 \si{km.s^{-1}}. Not only that we have at Jupiter a $\Delta V$ for perijove of 8.632 \si{km.s^{-1}}.\\

\emph{Results: Payload 860kg. Launch on 2031 MAR 01, Mission Duration $\sim{23}$ years. (The declination of the escape asymptote for the mission is -22.8\degree.)}\\

\textbf{Scenario 2 (3 stages):}
We shall again assume a payload mass of 860 kg. The three stages fired at perijove, in sequence are CASTOR 30XL, CASTOR 30B and then finally a STAR 48B. The total payload mass of all these stages corresponds to a $V_{\infty}$ of about 11.8 \si{km.s^{-1}}. The total $\Delta V$ at Jupiter is 9.829 \si{km.s^{-1}}.\\

\emph{Results: Payload 860kg. Launch on 2031 FEB 28, Mission Duration$\sim{20}$ years. (The declination of the escape asymptote for the mission is -22.8\degree , as above.)}\\

Note here that with two or more refuelled Starships in LEO, we have the capacity to send more than two spacecraft to 'Oumuamua and further that one Starship alone could also achieve multiple payloads to 'Oumuamua, although at the price of longer flight times.\\

\section{Long March 9}
This Chinese launcher is currently under development and will boast a capacity of 150 mt to LEO, with two methane + LOX stages. Its first human-rated launch will be in 2030. One would expect this mix of propellants would entail the same drawback as the SH + Starship, concerning its lack of performance for interplanetary missions.\\

Fortunately, not so. The reason is that it is planned to be able to carry an additional LOX/LH2 3rd stage, allowing it to undertake missions to Jupiter, and so it is more than adequate for Project Lyra. \\

I estimate a payload mass of at least 9.2 mt to Jupiter. If we adopt a STAR 63F followed then by a STAR 48B, with a payload mass of 100kg, this enables a total $\Delta V$ to be burned at Jupiter of 9.3 \si{km.s^{-1}}.\\

\emph{Results: Payload 100kg. Launch on 2030 JAN 13, Mission Duration = 36 years.}\\

It is re-iterated here that, because this launcher is at an early stage of development and also very little information has been released, this is only an estimate of its capability. Furthermore, in regard to the possibility of refuelling in LEO and what that could leverage (refer Section 5), it should be regarded as a deliberately conservative one.

\section{Discussion}

In this technical note we have analysed various different launcher options which will be available for a mission to ‘Oumuamua, ‘Project Lyra’. All trajectory options exploited a Jupiter encounter, either passive (Ariane 6 4) or a powered Jupiter Oberth Manoeuvre (Falcon Heavy, SH + Starship, Long March 9).\\

We found by far the most powerful performance is provided by a refuelled Starship in LEO, the refuelling entails the launch of eight additional Starships with a propellant payload.\\ 

Currently a SpaceX launch on a Falcon Heavy is around \$1500 per kg (\$95M per launch), and the projected cost of a Starship launch is as low as \$150 per kg (\$22M per launch). Note however that SpaceX owner Elon Musk has big plans to reduce launch costs on the order of \$1M per launch, giving for the scenarios mentioned here, with 9 Starship launches, a total launch cost of \$9M, also by far the cheapest launch option. \\

Falcon Heavy is the surprise because the combination of trajectory and solid booster stages considered hits a \emph{sweet spot} in terms of reaching ‘Oumuamua particularly propitiously, faring well in comparison to the much more powerful Starship and Long March 9 for example.

\section{Conclusion}

Various launchers have been investigated to determine their performance for a mission to 'Oumuamua (in terms of minimising flight duration).\\

Perhaps a surprise is the ability of much less powerful launchers previously not considered by Project Lyra, to nevertheless deliver a spacecraft to 'Oumuamua with flight times on the order of decades. The \emph{V$_{\infty}$ Leveraging Manoeuvre} (VILM) is the crucial trajectory option by which this performance can be delivered.\\

However the launcher which outshines all other candidates, due to its potential for LEO refuelling, is the SpaceX Super-Heavy + Starship, which would quite easily allow 'Oumuamua intercept in less than 20 years, and undoubtedly opens up the possibility of multiple spacecraft being deployed for Project Lyra.

\bibliographystyle{jasr-model5-names}
\biboptions{authoryear}
\bibliography{Which_Launchers,library_Adam_Hibberd,Hein_ISO_Modified_by_Adam_Hibberd}{}

\end{document}